\documentstyle[12pt,epsfig,psfig]{article}
\def\VEV#1{\left\langle #1\right\rangle}
\def\fig#1{{Fig. (\ref{#1})}}
\def\eq#1{{eq. (\ref{#1})}}

\def\gsim{\;
\raise0.3ex\hbox{$>$\kern-0.75em\raise-1.1ex\hbox{$\sim$}}\;}
\def\lsim{\;
\raise0.3ex\hbox{$<$\kern-0.75em\raise-1.1ex\hbox{$\sim$}}\;}

\newcommand\beq{\begin{equation}}
\newcommand\eeq{\end{equation}}

\newcommand\sdt{\sin^22\theta}

  \def\gsim{\;
  \raise0.3ex\hbox{$>$\kern-0.75em\raise-1.1ex\hbox{$\sim$}}\; }
\def\lsim{\;
  \raise0.3ex\hbox{$<$\kern-0.75em\raise-1.1ex\hbox{$\sim$}}\; }

\begin{document}
\thispagestyle{empty}
\begin{titlepage}
 \begin{flushright} \small \rm
   hep-ph/0005259 \\
   FTUV/000522 \\
   IFIC/00-26
 \end{flushright}
  \centerline{\bf \large The Simplest Resonant
   Spin--Flavour Solution to the Solar Neutrino Problem} \vskip 0.3cm
 \normalsize \vskip1cm
\begin{center}
  \baselineskip=13pt
  { O. G. Miranda~$^b$} \footnote{ E-mail: omr@fis.cinvestav.mx}
  { C. Pe\~na-Garay~$^a$}, \footnote{ E-mail:
    penya@flamenco.ific.uv.es}
  { T. I. Rashba~$^c$}, \footnote{ E-mail: rashba@izmiran.rssi.ru}
  { V. B. Semikoz~$^{a,c}$}, \footnote{ E-mail:
    semikoz@flamenco.ific.uv.es}\\ \vskip 0.3cm
  { and J.~W.~F. Valle~$^a$} \footnote{ E-mail:
    valle@flamenco.ific.uv.es}
\end{center}
\begin{center}
  {\it $^a$ Instituto de F\'{\i}sica Corpuscular -
    C.S.I.C./Universitat de Val\`encia}\\
  {\it Edificio Institutos de Paterna, Apartado de Correos 2085}\\
  {\it 46071 Val\`encia, SPAIN }\\
  {\it http://neutrinos.uv.es        }\\
\end{center}
\begin{center}
  {\it $^b$ Departamento de F\'{\i}sica \\ CINVESTAV-IPN, A. P.
    14-740, M\'exico 07000, D. F., M\'exico.} \\
\end{center}
\begin{center}
  {\it $^c$ The Institute of the Terrestrial Magnetism, \\
    the Ionosphere and Radio Wave Propagation of the Russian
    Academy of Science, \\
    IZMIRAN, Troitsk, Moscow region, 142092, Russia} \\
\end{center}

\baselineskip=13pt

\begin{abstract}
  
  We re-analyse the resonant spin--flavour (RSF) solutions to the
  solar neutrino problem in the framework of analytic solutions to the
  solar magneto-hydrodynamics (MHD) equations.
  By substantially eliminating the arbitrariness associated to the
  magnetic field profile due to both mathematical consistency and
  physical requirements we propose the simplest scheme (MHD-RSF, for
  short) for solar neutrino conversion using realistic static MHD
  solutions.
  Using such effective two--parameter scheme we perform the first
  global fit of the recent solar neutrino data, including event rates
  as well as zenith angle distributions and recoil electron spectra
  induced by solar neutrino interactions in Superkamiokande.
  We compare quantitatively our simplest MHD-RSF fit with vacuum
  oscillation (VAC) and MSW--type (SMA, LMA and LOW) solutions to the
  solar neutrino problem using a common well--calibrated theoretical
  calculation and fit procedure.
  We find our MHD-RSF fit to be somewhat better than those obtained
  for the favored neutrino oscillation solutions, though not in a
  statistically significant way with $\Delta m^2 \approx 10^{-8}
  eV^2$ and $\sdt =0$.
  We briefly discuss the prospects to disentangle our MHD-RSF scenario
  from oscillation--type solutions to the solar neutrino problem at
  future solar neutrino experiments, giving some predictions for
  the SNO experiment.\\
\end{abstract}
\end{titlepage}

\section{Introduction}

The persistent disagreement between solar neutrino data and
theoretical expectations has been a long-standing problem in physics.
Since the very first measurements \cite{homestake0}, the Solar
neutrino problem has remained as a puzzle, re-confirmed by new data on
rates by GALLEX-SAGE \cite{gallex,sage} as well as most recently
published 825--day data collected by the Super-Kamiokande
collaboration \cite{sk800} which goes beyond the simple rate
measurements to include also rate--independent data such as the recoil
electron spectra induced by solar neutrino interactions, as well as
the zenith angle distributions~\cite{sk99}.  It has often been argued
that these data can not be accounted for by astrophysics
\cite{Fiorentini:1997cf}.
Together with the atmospheric neutrino data \cite{Fukuda:1998mi} these
constitute the only present--day evidence in favour of physics beyond
the Standard Model, providing a strong hint for neutrino conversion.

The most popular solutions of the solar neutrino anomalies are based
on the idea of neutrino oscillations, either in vacuum or in the Sun
due to the enhancement arising from matter effects~\cite{MSW}.

Although these are the simplest neutrino conversion mechanisms there
is considerable interest in alternative interpretations. For example
it has long been noted \cite{Schechter:1981hw} that Majorana neutrinos
may have non--zero transition magnetic moments which can generate
spin--flavour conversions in the presence of a magnetic field. These
are especially interesting for two reasons: (i) on general grounds
\cite{Schechter:1980gr} neutrinos are expected to be Majorana
particles and (ii) conversions induced by transition magnetic moments
can be resonant in the Sun~\cite{Akhmedov}. There is also room for
more exotic mechanisms such as flavour changing neutrino interactions
\cite{Bergmann:2000gp} which do not even require neutrino mass
~\cite{Valle:1987gv}.

Here we will re-analyse the status of resonant spin--flavour solutions
to the solar neutrino problem in the light of the most recent global
set of solar neutrino data, including event rates as well as zenith
angle distributions and recoil electron spectra induced by solar
neutrino interactions in Superkamiokande which has attracted interest
recently~\cite{ad-hoc-profile1,ad-hoc-profile2, Morocco}.
In contrast to previous attempts we will adopt the general framework
of self--consistent magneto--hydrodynamic (MHD) models of the
Sun~\cite{Yoshimura}.  A previous attempt in this direction is given
in ref.\cite{Krastev:1991nr}. For definiteness we will concentrate in
the recent proposal of Ref.~\cite{Kutvitsky} where relatively simple
analytic solutions have been given.
We perform global fits of solar neutrino data for realistic solutions
to the magneto-hydrodynamics equations inside the Sun.  This requires
adjusting both the neutrino parameters as well as optimizing the
magnetic field profile. The arbitrariness associated to the latter is
substantially reduced due to mathematics (they must be solutions of
MHD equations) as well as reasonable physical requirements.
This way and by neglecting neutrino mixing we obtain the simplest
MHD-RSF solution to the solar neutrino problem, characterized by two
effective parameters, $\Delta m^2$ and $\mu_\nu B_{\perp max}$,
$B_{\perp max}$ being the maximum magnitude of the magnetic field
inside the convective region. Throughout this paper we have assumed
that the neutrino transition magnetic moment $\mu_{\nu}$ is given in
units of $\mu_{11} \equiv \mu_\nu / 10^{-11}$ $\mu_{B}$, where
$\mu_{B}$ is the Bohr magneton and we set $\mu_{11} \equiv 1$
everywhere.  Our MHD-RSF solution can be meaningfully compared with
the neutrino oscillation solutions to the solar neutrino problem.  We
find that our simplest two-parameter MHD-RSF fits to the solar
neutrino data are slightly better than those for the oscillation
solutions, but not in a statistically significant way.
The required best fit points correspond to maximum magnetic field
magnitudes in the convective zone smaller than 100 KG.
We briefly discuss the prospects to distinguish our simplest MHD-RSF
scenario from the neutrino oscillation solutions to the solar neutrino
problem at future solar neutrino experiments, giving some predictions
for the SNO experiment.

\section{Static Magnetic Field Profiles in the Sun}

In solar magneto-hydrodynamics~\cite{Parker} (MHD, for short) one can
explain the origin of solar magnetic fields from the dynamo mechanism
at the bottom of the convective zone or, to be more specific, in the
overshoot layer, where magnetic fields may be as strong as 300~kG
\cite{Bmax}.  Such a picture is quite attractive and several MHD
dynamo solutions has been known since long time ago (see for example
\cite{Yoshimura})
However the corresponding magnetic field profiles are rather
complicated and difficult to extract. For this reason there have been
many attempts to mimic MHD properties through the use of {\it ad hoc}
magnetic field profiles involving, for example, twisting fields
~\cite{Schechter}.

Here we will follow an alternative approach using fully
self-consistent solutions to the MHD equations inside the Sun. To
achieve this we focus on the case of stationary solutions which are
known analytically in terms of relatively simple functions
\cite{Kutvitsky}. This way we obtain a simple and well-motivated
magnetic field profile, without the full complexity that a dynamo
model implies.
In this section we will explain this model and discuss the limits on
the shape parameters describing the field profile. We will also
discuss how to relate this model with the dynamo picture of the solar
interior.

\subsection{Single-Mode  Field Configurations}

In this subsection we will describe the model we are using for the
magnetic field profile.  We consider only solutions to the equation
for a static MHD plasma configuration in a gravitational field given
by the equilibrium of the pressure force, the Lorentz force and the
gravitational force
\begin{equation}
\label{maxwell}
\nabla p -\frac1c \vec{j} \times \vec{B} + \rho \nabla \Phi =0 ,
\end{equation}
where p is the pressure, $j = (c/4\pi)rot~\vec{B}$ is the electric
current, $B$ is the static magnetic field under consideration, $\rho$
is the matter density \cite{Kutvitsky} and $\Phi$ is the gravitational
potential.

This static MHD equations correspond to a quiet Sun and they admit
axially symmetric solutions in the spherically symmetric gravitational
field which can be simply expressed in terms of spherical Bessel
functions and were first discussed in Ref.  \cite{Kutvitsky}.
For this model the magnetic field will be given by a family of
solutions that depends on $z_{k}$, the roots of the spherical Bessel
function $f_{5/2}=\sqrt{z}J_{5/2}(z)$, to ensure the boundary
condition that $\vec{B}$ vanishes on the solar surface. Within the
solar interior the magnetic field for any $k$ will be then given by
\begin{eqnarray}
B^k_r (r, \theta) &=& 2\hat B^k\cos \theta\left[1 -
\frac{3}{r^2z_{k}\sin z_{k}}
\left(\frac{\sin(z_{k}r)}{z_{k}r} - \cos(z_{k}r)\right)\right]~,
\nonumber\\
B^k_{\theta} (r, \theta) &=& - \hat B^k\sin \theta\left[2  +
\frac{3}{r^2z_{k}\sin z_{k}}\left(\frac{\sin(z_{k}r)}{z_{k}r} - 
\cos(z_{k}r) - z_{k}r\sin(z_{k}r)\right)\right]~,\nonumber\\
B^k_{\phi} (r, \theta) &=& \hat B^k z_{k}\sin \theta\left[r -
\frac{3}{rz_{k}\sin z_{k}}\left(\frac{\sin(z_{k}r)}{z_{k}r} - 
\cos(z_{k}r)\right)\right]~,
\label{kutv}
\end{eqnarray}
where the coefficient $\hat B^k(B_{core})$ is given by
\begin{equation}
\hat B^k = \frac{B_{core}}{2(1 - z_{k}/\sin z_{k})}~.
\end{equation}
Here $\theta$ is the polar angle and the distance $r$ has been
normalized to $R_\odot=1$.  Taking into account the inclination of the
solar equator to the ecliptics, where neutrinos propagate to the
Earth, it follows that $\theta$ lies in the narrow range $83^o-97^o$,
depending on the season. In our calculations we have averaged over
$\theta$ in the above range.

The modulus of the perpendicular component which is relevant to the
neutrino spin-flavour takes the form
\begin{equation}
B_{\perp} = \sqrt{B_{\phi}^2 + B_{\theta}^2} = B_{core}\frac{\sin \theta}{r}f(r)~,
\label{perp}
\end{equation}
where $f(r)$ is some known smooth function.  Notice also that the
behaviour of {\bf B} at the solar center (r = 0)
\begin{eqnarray}
B_r (0,\theta) &=& B_{core}\cos \theta~,\nonumber\\
B_{\theta}(0,\theta) &=& - B_{core}\sin \theta~,\nonumber\\
B_{\phi}(0,\theta) &=& B_{core}\sin \theta\frac{z_{k}}{2}\frac{r}{R_{\odot}}
\to 0~,
\label{zero}
\end{eqnarray}
is completely regular, determined only by the parameter $B_{core}$. In
Fig. 1 we display the perpendicular component of {\bf B} for various
$k$--values 1, 3 and 10, which correspond to the roots $z_k=5.7$,
$z_k=12.3$ and $z_k=34.5$, respectively.
\begin{figure}
  \centerline{\protect\hbox{\psfig
      {file=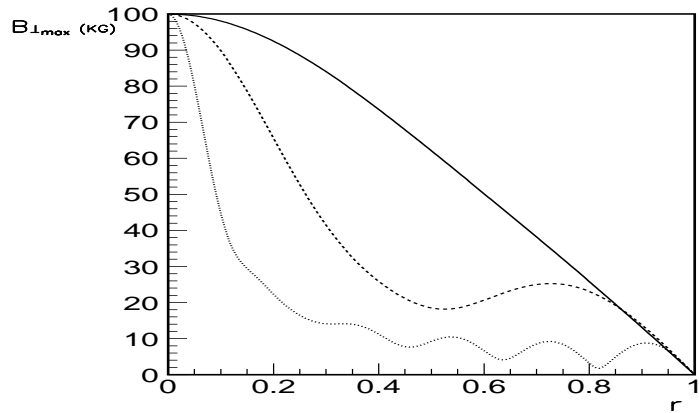,height=6cm,width=10cm,angle=0}}}
\caption{The perpendicular component of {\bf B} for various
  $k$--values 1 (solid), 3 (dashed) and 10 (dotted). }
\end{figure}

\subsection{Astrophysical Constraints on Magnetic Fields }

We now discuss the astrophysical restrictions on the free parameters
$B_{core}$ and $k$ characterizing the model. We can see that the
magnitude of a magnetic field at the center of the Sun is constrained
by the Fermi-Chandrasekhar limit~\cite{Chandra} which implies
\cite{Kutvitsky}
\begin{equation}
\eta=\frac{2}{15}\frac{5\gamma_0-6}{\gamma_0-1}
\frac{(\hat B^k)^2 z^2_{2k}R_\odot^4}{GM_\odot^2} \lsim 1,
\label{Chandrasekhar}
\end{equation}
here $\gamma_0$ is polytropic index characterizing the equation of
state (pressure $P \sim \rho^{\gamma_0}$), $M_\odot$ is the solar mass
and $G$ is Newton's constant. This equation gives us an upper bound on
$B_{core} \lsim 2~$ MGauss for $k=1$. For higher values of $k$ this
constraint is even weaker.

Regarding with the values of $k$. These can be constrained by taking
into account that in order to justify the use of a stationary
solution, it is necessary that the diffusion time due to ohmic
dissipation
\begin{equation}
\label{diss}
t_{diss}(k)=\frac{4\pi\sigma_{cond}L^{2}(k)}{c^{2}}
\end{equation}
must be bigger than the age of the Sun $t_{\odot} \simeq 1.4 \times
10^{17}s$ \cite{SunAge}. Here $L(k)$ denotes a characteristic spatial
scale of the magnetic field which corresponds to the typical distance
between subsequent nodes of the corresponding Bessel function. As we
can see from Fig. 1 $L(k=1) \sim R_{\odot}$, while $L(k) \simeq
R_{\odot}/k$. In \eq{diss}
$\sigma_{cond}=\omega_{p}^{2}/(4\pi\nu_{ep})$ denotes the conductivity
of the fully ionized hydrogen plasma. After substituting the plasma
frequency
\[
\omega_{p}=5.65\times10^{4}\sqrt{\frac{n_{e}}{1cm^{-3}}}~s^{-1}
\]
and the $e$-$p$ collision frequency
\[
\nu_{ep}=50\left( \frac{n_{e}}{1cm^{-3}}\right) \left(
  \frac{T}{1K}\right) ^{-3/2}~s^{-1}
\]
we obtain from \eq{diss} an estimate of the magnetic field dissipation
time $t_{diss}(k)$
\begin{equation}
t_{diss}(k)=\frac{6.4\times 10^{7}
R^{2}_\odot1/s}{c^{2}}\times \frac{(T/K)^{3/2}}{k^{2}}>t_{\odot}
\end{equation}
Note that the dissipation time is shorter for higher $k$ values,
long--lived field configurations being possible only if $k$ is small.
For example, the dissipation time for the third mode is about an order
magnitude less than that for the first mode.  Moreover it depends on
the value of the temperature that we take.  Some typical values for
the temperature are $T_{\min} \simeq 2.8 \times10^{5} K$ for the
bottom of the convective zone and $T_{\max}\simeq1.6\times10^{7}K$ for
the solar core \cite{BP98}.  Thus, taking the optimistic estimate,
$T_{\max}$, we obtain $k<k_M=13$, while, if we consider the average
value $T=4\times10^{6}K$ we will have $k<k_M=5$. In what follows we
will consider values of $k \leq 10$.

\subsection{Energy localization criterium}

It is commonly accepted that magnetic fields measured at the surface
of the Sun are weaker than within the convective zone interior where
this field is supposed to be generated. It is known by observational
data that the mean field value over the solar disk is of the order of
1 Gauss while in the solar spots magnetic field strength reaches 1 KG.

On the other hand the general knowledge of the solar magnetic field
models is that the magnetic field increases at the overshoot layer,
while being small at the solar interior, a picture rather opposite to
the one we have seen in Fig. 1.

Although there is no direct information on the magnetic field
magnitudes at the solar core, there are theoretical reasons which
imply a central magnetic field less than 30 Gauss~\cite{boruta};
otherwise the present magnetic field in the convective zone would be
too big, leading to a visible enhancement in sunspot activity.

This conflict can be avoided by taking advantage of the linear nature
of the basic equilibrium MHD equation in \eq{maxwell}. This implies
that any linear combination of solutions $\vec{B}^k$ ($k=1, 2, \dots,
k_M$, for some fixed number $k_M$)
\begin{equation}
\vec{B}=c_1\vec{B_1}+c_2\vec{B_2}+...+c_M\vec{B_M}
\label{combine}
\end{equation}
is also a solution.  As mentioned in section 2.2 we will adopt $k_M
\leq 10$ in order to ensure that ohmic dissipation is acceptable and
therefore justify the static approximation. 
\begin{figure}
  \centerline{\protect\hbox{\psfig
      {file=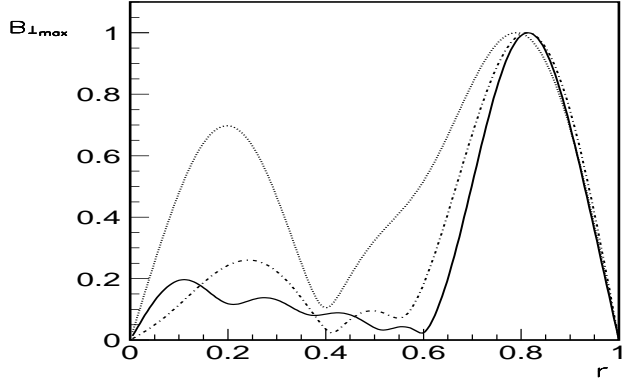,height=6cm,width=10cm,angle=0}}}
\caption{Magnetic field configurations obtained by combining individual 
  modes for different $k_M$ values, 5, 6 and 10. Summing up to higher
  modes achieves better localization of the field in the convective
  region (solid). }
\label{combi}
\end{figure}

In order to ensure that the magnetic field energy is localized mainly
within the convective region we will now supplement the constraints of
section 2.2 by imposing that the magnetic field should vanish in the
center of the Sun
\footnote{ Note that adopting a finite but small value for
  $|\vec{B}(\vec{r}=0)| \lsim 30$~Gauss does not change significantly
  the profile of the total magnetic field \eq{combine} which is the
  relevant quantity for describing neutrino propagation.}
\begin{equation}
\vec{B}(\vec{r}=0)=0
\end{equation}

The latter implies
\begin{equation}
c_1+c_2+...+c_{k_M}=0.
\label{restriction1}
\end{equation}
Therefore we will have, in principle, ${k_M}-1$ free parameters.

We will require, in addition, that the magnetic field energy must be
minimal in the region below the bottom of the convective zone,
characterized by a certain value of $r_0$
\begin{equation}
E_B = \int_0^{r_{0}}d^3r\frac{{\vec B}^2(r)}{8\pi}~.
\label{restriction2}
\end{equation}
This implies
\begin{equation}
\frac{\partial}{\partial c_i} E_B \equiv \Sigma^{{k_M}-1}_{j=1}c_j a_{ij}=0
\end{equation}
where
\begin{equation}
a_{ij}=\int_0^{\vec{r_{0}}}d^3r(\vec{B^j}-\vec{B^M})\cdot\vec{B^i}~.
\end{equation}

Without loss of generality we can assume that one of the coefficients
is non-zero, which prevents us from having only the trivial solution
$c_i \equiv 0$. Taking $c_{{k_M}-1} \neq 0$ we will have the linear
non--homogeneous equation
\begin{equation}
\Sigma^{M-2}_{j=1}c_j a_{ij}=-c_{{k_M}-1} a_{_{i{k_M}-1}}
\end{equation}
which determines all the $c_i$ coefficients in terms of, say,
$c_{{k_M}-1}$. As expected on physical grounds, this last remaining
parameter $c_{{k_M}-1}$ corresponds to the maximum magnetic field
magnitude in the convective region, i.e.  $B_{\perp_{max}}$ is
proportional to $c_{{k_M}-1}$.  In general, all the coeficients $c_i$
will be different from zero for $i<M$ and alternate in sign. As an
example for $k_M=6$ the coefficients are -0.338968, -0.261825,
1.29186, -1.77360, 1.86916, -0.786628.

The procedure sketched above provides a consistent method for
combining individual mode solutions $\vec{B_k}$ of the static MHD
equation, while fixing all of the coefficients of the linear
combination, leaving as free parameters only the value of
$B_{\perp_{max}}$ inside the convective region and the value of $k_M
\leq 10$.  In \fig{combi} we show the resulting combined profiles for
$k_M=5, 6, 10$. The parameter $r_0$ could also be taken as a free
parameter but, on physical grounds, it should lie in a narrow range
close to overshoot layer. We show explicitly that varying $r_0$ has
little effect on our results.

\section{ Fitting the Solar Neutrino Data }

The Majorana neutrino evolution Hamiltonian in a magnetic field is
well--known to be four--dimensional~\cite{Schechter:1981hw}.
For definiteness and simplicity we will neglect neutrino mixing in
what follows and consider first the case of active-active neutrino
conversions. This will allow us to compare our $\chi^2$-analysis with
the previous ones \cite{ad-hoc-profile1,ad-hoc-profile2,Morocco}. The
$\nu_{e}\to \bar{\nu}_{\ell}$ conversions are described by the master
Schr\"{o}dinger evolution equation
\begin{equation}
i\left(
\begin{array}{l}
\dot{\nu}_{e}\\
\dot{\bar{\nu}_{\ell} }\\
\end{array}
\right) =
\left(
\begin{array}{cc}
V_e - \delta &  \mu_\nu B_{+}  \\
\mu_\nu B_{-} & - V_{\ell} + \delta  \\
\end{array}
\right)
\left(
\begin{array}{c}
\nu_{e}\\
\bar{\nu}_{\ell} \\
\end{array}
\right)~,
\label{master}
\end{equation}
where $\mu_\nu$ denotes the neutrino transition magnetic
moment~\cite{Schechter:1981hw} in units of $10^{-11}$ $\mu_B$, $\ell$
denoting either $\mu$ or $\tau$. Here $B_\pm=B_x\pm iB_y$ and $\delta
= \Delta m^2/4E$ is the neutrino mass parameter; $V_e(t)
=G_F\sqrt{2}(\rho (t)/m_p)(Y_e - Y_n/2)$ and $V_{\ell}(t)
=G_F\sqrt{2}(\rho (t)/m_p)(- Y_n/2)$ are the neutrino vector
potentials for $\nu_{e}$ and $\nu_{\ell}$ in the Sun given by the
abundances of the electron ($Y_e = m_p N_e(t)/\rho (t)$) and neutron
($Y_n =m_pN_n(t)/\rho (t)$) components. In our numerical study of
solar neutrino data we adopt the Standard Solar Model density profile
of ref.{\cite{BP98}.
  
We solve Eq. (\ref{master}) numerically by finding a solution of the
Cauchy problem in the form of a set of wave functions $\nu_a(t)= \mid
\nu_a (t)\mid e^{i\Phi_a (t)}$ from which the neutrino survival
probabilities $P_{aa}(t)= \nu_a^*\nu_a$ are calculated.  They obey the
unitarity condition $\sum_a P_{aa} = 1$ where the subscript $a$
denotes $a=e$ for $\nu_{e}$ and $a = \ell$ for $\bar{\nu}_{\ell }$
respectively.
  
As an illustration we display in \fig{MHD-RSF-prob} the electron
neutrino survival probablity $P_{ee}$ calculated in the MHD-RSF scheme
from \eq{master} plotted versus $E/\Delta m^2$. This is obtained with
the magnetic field configurations given in \fig{combi}.
\begin{figure}
\centerline{\protect\hbox{\psfig{file=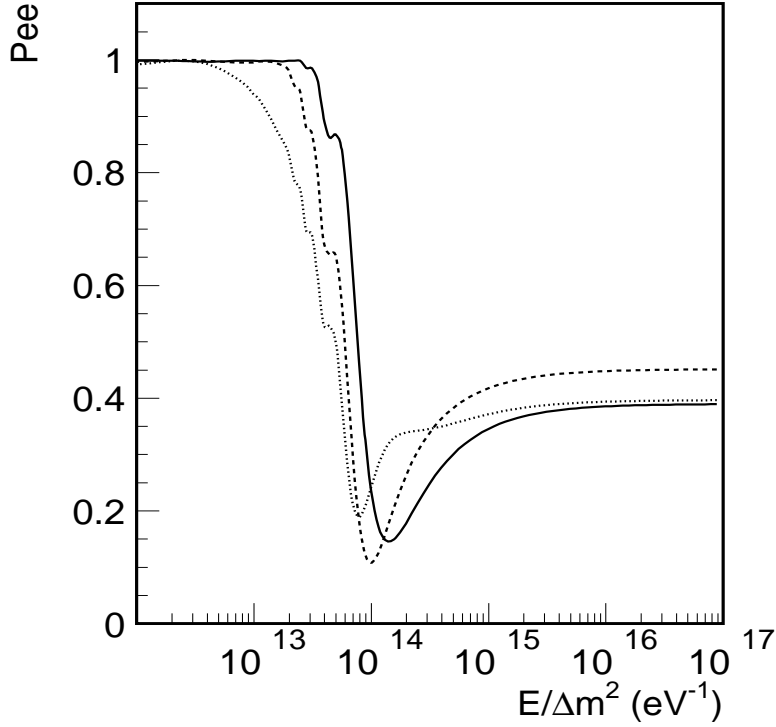,height=11cm,width=12cm,angle=0}}}
\caption{Typical MHD-RSF neutrino survival probablity $P_{ee}$ 
versus $E/\Delta m^2$. }
\label{MHD-RSF-prob}
\end{figure}

We will now re-analyse the status of resonant spin--flavour solutions
to the solar neutrino problem in the light of the most recent global
set of solar neutrino data, including event rates as well as zenith
angle distributions and recoil electron spectra induced by solar
neutrino interactions in Superkamiokande which has attracted interest
recently~\cite{ad-hoc-profile1,ad-hoc-profile2,Morocco}.  It has been
found that the quality of the fit to the solar neutrino data depends
on the magnetic field profile. The best solutions have been obtained
with a magnetic field around 100 KG in the convective zone and zero at
the core (profile 3 in Ref~\cite{ad-hoc-profile1}, and profile 6 of
Ref~\cite{ad-hoc-profile2}) or an almost constant magnitude of the
magnetic field, but with twisting direction \cite{Morocco} using the
profiles given in \cite{Schechter}.
  
  In contrast with previous work we will consider the fits obtained
  when we employ self--consistent solutions of MHD equations which
  obey the physical requirements we derived in sec. 2.2 using the
  procedure for combining magnetic field modes described in section 3.
  Our approach is \emph{global} and allows us to compare quantitatively
  with other solutions to the solar neutrino problem within the same
  well calibrated theoretical calculation and fit procedure.

\subsection{Rates}

In order to determine the possible values of the parameters
characterizing the MHD-RSF solution to the solar neutrino problem, we
have first used the data on the total event rates measured at the
Chlorine experiment in Homestake \cite{homestake0}, at the two Gallium
experiments GALLEX and SAGE \cite{gallex,sage} and the 825-day
Super--Kamiokande data sample, as given in table~\ref{rates12}.
\begin{table}
\begin{center}
\begin{tabular}{|l|l|l|l|l|}
\hline
Experiment & Rate & Ref. & Units& $ R^{\rm BP98}_i $\\
\hline
Homestake  & $2.56\pm 0.23 $ & \protect\cite{homestake0} & SNU &  $7.8\pm 1.1 $   \\
GALLEX + SAGE  & $72.3\pm 5.6 $ & \protect\cite{gallex,sage} & SNU & $130\pm 7 $  \\
Super--Kamiokande & $2.45\pm 0.08$ & \protect\cite{sk800} & 
$10^{6}$~cm$^-2$~s$^{-1}$ & $5.2\pm 0.9 $ \\   \hline
\end{tabular}
\vglue .3cm 
\caption{Solar neutrino rates measured in the Chlorine, Gallium 
and Super--Kamiokande experiments. }
\label{rates12}
\end{center}
\end{table}

In our statistical treatment of the data for the combined fit we adopt
the $\chi^2$ definition given in ref.~\cite{two},
\begin{equation}
\chi^2_R=\sum_{i,j=1,3} (R^{th}_i- R^{exp}_i)~,
\sigma_{ij}^{-2} (R^{th}_j- R^{exp}_j)
\end{equation}
where $R^{th}_i$ is the theoretical prediction of the event rate in
detector $i$ and $R^{exp}_i$ is the measured rate. The error matrix
$\sigma_{ij}$ contains not only the theoretical uncertainties but also
the experimental errors, both systematic and statistical.

The general expression of the expected event rate in the presence of
oscillations in experiment $i$ is given by $R^{th}_i$
\begin{eqnarray}
R^{th}_i & = & \sum_{k=1,8} \phi_k
\int\! dE_\nu\, \lambda_k (E_\nu) \times 
\big[ \sigma_{e,i}(E_\nu)  \VEV{P_{ee} (E_\nu,t)}  \label{ratesth} \\
& &                            + \sigma_{x,i}(E_\nu) 
(1-\VEV{P_{ee} (E_\nu,t)} )\big] \nonumber,
\end{eqnarray}  
where $E_\nu$ is the neutrino energy, $\phi_k$ is the total neutrino
flux and $\lambda_k$ is the neutrino energy spectrum (normalized to 1)
from the solar nuclear reaction $k$ ~\cite{Bspe} with the
normalization given in Ref.~\cite{BP98}. Here $\sigma_{e,i}$
($\sigma_{x,i}$) is the $\nu_e$ ($\nu_x$) (with $x$ being $\bar\mu$ or
$s$ corresponding to active-active or active-sterile MHD-RSF
conversions) cross section in the Standard Model~\cite{CrSe} with the
target corresponding to experiment $i$, and $\VEV{ P_{ee} (E_\nu,t) }$
is the time--averaged $\nu_e$ survival probability.

For the Chlorine and Gallium experiments we use improved cross
sections $\sigma_{i}(E)$ from Ref.~\cite{prod}. For the
Super--Kamiokande experiment we calculate the expected signal with the
corrected cross section given in the Appendix Sec.~\ref{dataspec}.

The expected signal in the absence of oscillations, $R^{\rm BP98}_i$,
can be obtained from Eq.(\ref{ratesth}) by substituting $P_{ee}=1$. In
table~\ref{rates12} we also give the expected rates at the different
experiments which we obtain using the fluxes of Ref.~\cite{BP98}.

In \fig{figrates} we display the region of MHD-RSF parameters allowed
by the solar neutrino rates.
\begin{figure}
\centerline{\protect\hbox{\psfig{file=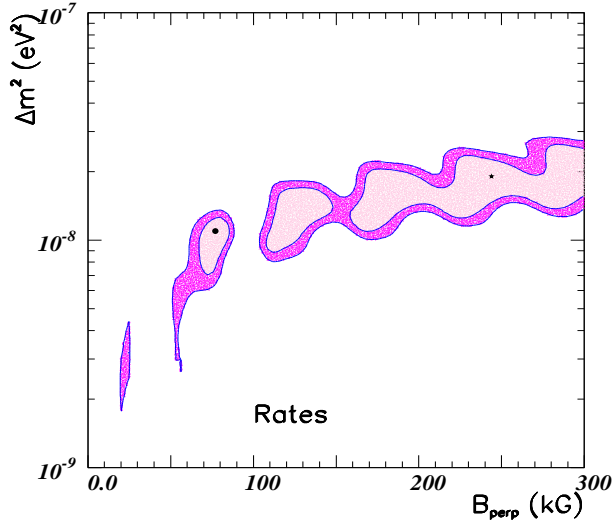,height=11cm,width=12cm,angle=0}}}
\vglue -3.5cm
\caption{MHD-RSF 90\% CL (light) and 99\% CL (dark) 
  regions of $\Delta m^2$ versus $B_{\perp max} (KG)$ allowed by the
  rates given in table~\protect\ref{rates12}, for $r_0 = 0.6$ and $K_M
  =6$ }
\label{figrates}
\end{figure}

Our $\chi^2$ analysis of the solar neutrino rates uses the magnetic
field profiles discussed in Section 2. As we mentioned in that
section, these profiles are characterized by $k_M$ and $r_0$. In table
\ref{fitrates} we present the best fit points for $k_M$ from 4 to 8
and for $r_0\simeq.6R\odot$ and for $B_{\perp_{max}}<300~$KGauss.  In
the same table we also show the best fit points for
$B_{\perp_{max}}<100~$KGauss.  We can see from this table that the
$\chi^2$ is pretty stable and does not depend significantly on the
choice of $k_M$ and $r_0$ allowed by astrophysics.  In \fig{figrates}
we display the region of MHD-RSF parameters allowed by the solar
neutrino rates for the case $M=6$ and $r_0=.6 R\odot$. We can see that
there are several allowed regions for different values of the magnetic
field. As we already mentioned, in our analysis we have fixed the
value of $\mu_{\nu}$ to be $10^{-11}\mu_B$. Since the evolution
equation depends on the product $\mu_{\nu}B$, for a smaller value of
the neutrino magnetic moment the $B_{\perp_{max}}$ in \fig{figrates}
would have to be correspondingly increased. In this sense,
the local minima shown in table~\ref{fitrates} for
$B_{\perp_{max}}<100~$ KGauss allows a smaller $\mu_{\nu}$.
\begin{table}
\begin{center}
\begin{tabular}{|c|c|c|c|c|c|c|c|c|c|}
\hline
 &&$0.6R_{\odot}$ &&& $0.62R_{\odot}$&&& $0.64R_{\odot}$&\\
\hline
$M$ &$B_{\perp {max}}$&$\Delta m^2$&$\chi^2_{min}$&$B_{\perp {max}}$&$\Delta m^2$& $\chi^2_{min}$&$B_{\perp {max}}$&$\Delta m^2$ & $\chi^2_{min}$ \\
\hline
4 &29.& $.84\times 10^{-8}$ &1.3& 32  &$.94\times 10^{-8}$ &1.1& 41 
&$.99\times 10^{-8}$&1.3\\
  &29.& $.84\times 10^{-8}$ &1.3& 32  &$.94\times 10^{-8}$ &1.1& 41 
&$.99\times 10^{-8}$&1.3\\
\hline
5 & 71 & $1.5\times 10^{-8}$&1.0 & 63& $1.5\times 10^{-8}$ &.87& 72& 
$1.6\times 10^{-8}$&.95\\
  & 71 & $1.5\times 10^{-8}$&1.0 & 63& $1.5\times 10^{-8}$ &.87& 72& 
$1.6\times 10^{-8}$&.95\\
\hline
6 & 244 & $1.9\times 10^{-8}$ &.03& 247& $1.9\times 10^{-8}$&.06&251& 
$1.9\times 10^{-8}$&.13\\
  & 77 & $1.1\times 10^{-8}$ &.33& 80& $1.0\times 10^{-8}$&.40&80& $1.0\times 
10^{-8}$&.35\\
\hline
7 & 208 & $1.2\times 10^{-8}$&.20 & 210 &$1.0\times 10^{-8}$&.13& 215 
&$.94\times 10^{-8}$&.26 \\
  & 83 & $.75\times 10^{-8}$&.52 & 84 &$.71\times 10^{-8}$&.54& 84 
&$.64\times 10^{-8}$&.47 \\
\hline
8 & 220 & $.98\times 10^{-8}$&.38 & 222& $.94\times 10^{-8}$&.24 &225 
&$.84\times 10^{-8}$&.45 \\
  & 87 & $.64\times 10^{-8}$&.68 & 87& $.59\times 10^{-8}$&.64 &87 
&$.55\times 10^{-8}$&.69 \\
\hline
\end{tabular}
\vglue .3cm 
\caption{Best fit points for the rates--only analysis for different $r_0$ and 
  $k_M$ values in active-active MHD-RSF oscillations. }
\label{fitrates}
\end{center}
\end{table}

\subsection{Zenith and Spectrum Fit}

Apart from total event rates the water Cerenkov experiment also
measures the zenith angle distribution of solar neutrino events as
well as their electron recoil energy spectrum with their recent
825-day data sample~\cite{sk800}.

The smallness of the $\Delta m^2$ values indicated by the rates fit
implies that no appreciable day--night variation of the counting rates
is expected in our MHD-RSF solution. However the measured solar
neutrino zenith angle data must be included in the analysis and we do
that.  This is necessary in order to enable us a meaningful comparison
with vacuum and matter oscillations using the same statistical
criteria~\cite{two,dark,four}, see definitions in the appendix.  We
obtain $\chi_{zenith}^2 = 5.4$ for the full range of parameters in the
analysis, the same as for the no-oscillation case.

The recoil electron energy spectrum induced by solar neutrino
interactions after 504 days of operation is given for energies above
6.5 MeV using the Low Energy (LE) analysis in which the recoil energy
spectrum is divided into 16 bins, 15 bins of 0.5 MeV energy width and
the last bin containing all events with energy in the range 14 MeV to
20 MeV.  Below 6.5 MeV the background of the LE analysis increases
very fast as the energy decreases.  Super--Kamiokande has designed a
new Super Low Energy (SLE) analysis in order to reject this background
more efficiently so as to be able to lower their threshold down to 5.5
MeV.  In their 825-day data \cite{sk800} they have used the SLE method
and they present results for two additional bins with energies between
5.5 MeV and 6.5 MeV.  In the appendix we present these data in
table~\ref{spectrum} as well as the details of our statistical
analysis.  Our results are almost independent of the choice of the
parameters $k_M$ and $r_0$ in the physical range, thus establishing
the robustness of the fit procedure. The predicted spectrum is
essentially flat except for the upper part of the $\Delta m^2$ region.
As an example, we show in fig.~ \ref{figspectrum} the excluded region
at 99 \% CL for the case $k_M=6$ and $r_0=0.6$.

\begin{figure}
  \centerline{\protect\hbox{\psfig
      {file=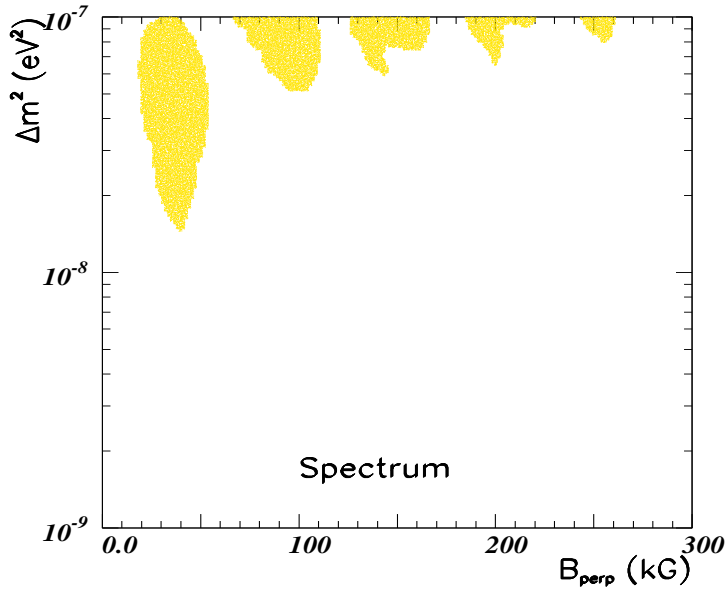,height=8cm,width=13cm,angle=0}}}
\caption{MHD-RSF 99\% CL regions of  
  $\Delta m^2$ versus $B_{\perp max}$ forbidden by the recoil electron
  spectrum data given in table \protect\ref{spectrum} of the appendix,
  for $r_0 = 0.6$ and $k_M =6$}
\label{figspectrum}
\end{figure}

\subsection{Global Fit}

As we have seen in the partial analysis, zenith and spectrum are
essentially flat in the region of parameters which provide a good fit
for the rates--only analysis. For this reason, the allowed regions are
slightly modified by the inclusion of the zenith angular dependence
and the energy spectrum data. As the results are statistically
independent of the choice of $k_M$ and $r_0$ in the physical range,
our analysis effectively involves only two parameters. It is therefore
meaningful to compare it with the popular two--neutrino fits
characterizing vacuum or matter--enhanced oscillations.  In
table~\ref{global}, we show the best--fit points in the range of our
study for different $k_M$ and $r_0$ values.  Moreover, we show the
local (global) minimum for $B_{\perp {max}}$ less than 100 KG, which
will be important to improve sensitivity on the transition magnetic
moment of the neutrino. In fig.~\ref{figspectrum} we show the allowed
region at 90\% CL and 99\% CL for the case $r_0 = 0.6$ and $k_M =6$.
We have also investigated the effect of varying the hep flux,
obtaining for the allowed regions results similar to the
no--oscillation solution discussed previously in ref.~\cite{two},
independently of the $\Delta m^2$ and $B_{\perp max}$ value, with a
$hep$ normalization factor of 13.5.
\begin{table}
\begin{center}
\begin{tabular}{|c|c|c|c|c|c|c|c|c|}
\hline
 & &$0.6R_{\odot}$ && & & $0.62R_{\odot}$& &\\
\hline
$M$ &$B_{\perp {max}}$&$\Delta m^2$&$\chi^2_{min}$ & $\alpha_{sp}$ &$B_{\perp {max}}$&$\Delta m^2$& $\chi^2_{min}$ & $\alpha_{sp}$ \\
\hline
5 & 72 & $1.6\times 10^{-8}$&25.7 & 1.10 & 72.& $2.0\times 10^{-8}$ &25.7 & 1.17 \\
  & 72 & $1.6\times 10^{-8}$&25.7 & 1.10 & 72.& $2.0\times 10^{-8}$ &25.7 & 1.17\\
\hline
6 & 240 & $1.8\times 10^{-8}$ &24.7& 0.97 & 241& $1.8\times 10^{-8}$&24.9 & 0.96 \\
  & 80 & $1.1\times 10^{-8}$ &25.7& 1.11 & 80& $1.1\times 10^{-8}$&25.9 & 1.10\\
\hline
\end{tabular}
\vglue .3cm 
\caption{Global best fit points and local minima for 
  $B_{\perp {max}}<100$ kGauss for different $r_0$ and $k_M$ values in
  active-active MHD-RSF conversion scenario. }
\label{global}
\end{center}
\end{table}
\begin{figure}
  \centerline{\protect\hbox{\psfig
      {file=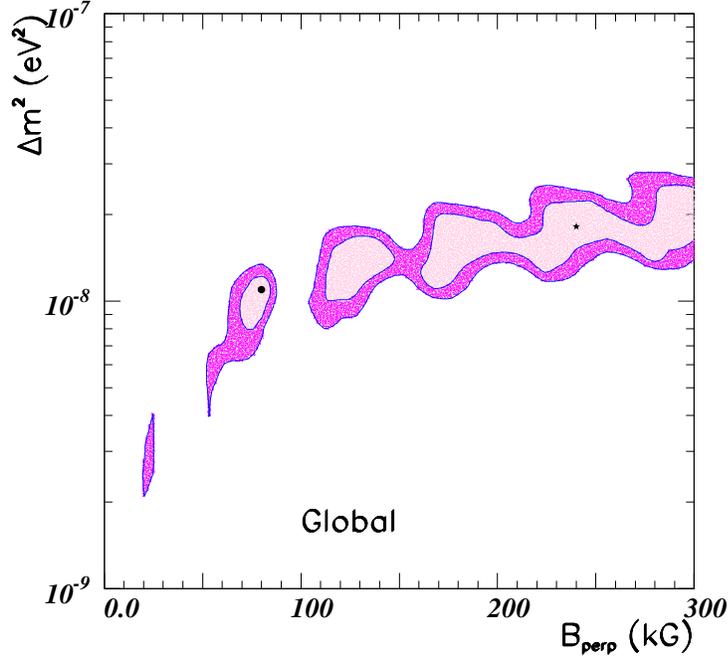,height=9cm,width=13cm,angle=0}}}
\caption{90\% CL (light) and 99\% CL (dark) allowed MHD-RSF regions in 
  $\Delta m^2$ and $B_{\perp max}$ from the measurements of rates
  combined with the zenith angle distribution and the recoil energy
  spectrum in super--Kamiokande, for $r_0 = 0.6$ and $k_M =6$. }
\label{figglobal}
\end{figure}


We now move to the case of active-sterile MHD-RSF conversions.  For
this case one must make substitute $\nu_s$ for $\nu_{\ell}$ in
\eq{master} and take into account that $V_s=0$.  The results given
above for active--active MHD-RSF conversions change when conversions
involve sterile neutrinos.  The best fit points ( and local ones ) are
obtained with parameters slightly modified with respect to those
obtained for the active--active case.  In the rates only fit, the
$\chi_{rates}^2$ is worse than for the active-active case, essentially
due to the neutral current contribution in the Super--Kamiokande
experiment. The zenith angle dependence and the recoil energy spectrum
remains flat as before. The global fit for different $r_0$ and $k_M$,
is shown in table~\ref{fitsterile}.
\begin{table}
\begin{center}
\begin{tabular}{|c|c|c|c|c|c|c|c|c|}
\hline
 &  &$0.6R_{\odot}$ &&&& $0.62R_{\odot}$&&\\
\hline
$M$ &$B_{\perp {max}}$&$\Delta m^2$&$\chi^2_{min}$&$\alpha_{sp}$&$B_{\perp {max}}$&$\Delta m^2$& $\chi^2_{min}$&$\alpha_{sp}$\\
\hline
5 &68.& $1.4\times 10^{-8}$ &34.0& 1.25&68  &$1.5\times 10^{-8}$ &35.2&1.30\\
  &68.& $1.4\times 10^{-8}$ &34.0& 1.25&68  &$1.5\times 10^{-8}$ &35.2&1.30\\
\hline
6  & 247. & $1.9\times 10^{-8}$&28.7 & 1.20&247.& $1.9\times 10^{-8}$ &28.2&1.18\\
  & 75. & $1.1\times 10^{-8}$&30.5 &1.13& 75.& $1.0\times 10^{-8}$ &30.6&1.10\\
\hline
\end{tabular}
\vglue .3cm 
\caption{Best fit points and local minima for $B_{\perp {max}}<100 kG$ for 
the global analysis for different $r_0$ and $k_M$ values in active-sterile 
MHD-RSF conversions. }
\label{fitsterile}
\end{center}
\end{table}

\section{MHD-RSF versus Oscillation Solutions}

\subsection{Present}

From the results of the previous section it follows that our MHD-RSF
solution to the solar neutrino problem provides a good description of
the most recent solar neutrino data, including event rates as well as
zenith angle distributions and recoil electron spectra induced by
solar neutrino interactions in Superkamiokande.  We have shown that
our procedure is quite robust in the sense that the magnetic field
profile has been determined in an essentially unique way. This
effectively substitutes the neutrino mixing which characterizes the
oscillation solutions by a single parameter $B_{\perp max}$
characterizing the maximum magnitude of the magnetic field inside the
convective region. The value of $k_M$ characterizing the maximum
number of individual modes superimposed in order to obtain a realistic
profile and the parameter $r_0$ characterizing the location of the
convective region are severely restricted. The allowed $k_M$ values
are restricted by ohmic dissipation arguments to be lower than 10 or
so, while $r_0$ is close to $0.6R_{\odot}$. We have found that our
solar neutrino fits are pretty stable as long as $k_M$ exceeds 5 and
$r_0$ lies in the relevant narrow range (see tables \ref{fitrates} and
\ref{global}).  Therefore our fits are effectively two--parameter fits
($\Delta m^2$ and $B_{\perp max}$) whose quality can be meaningfully
compared with that of the fits obtained for the favored neutrino
oscillation solutions to the solar neutrino problem.
In table~\ref{comp} we compare the various solutions of the solar
neutrino problem with the MHD-RSF solutions for the lower magnetic field 
presented here.
\begin{table} 
\begin{center}
\begin{tabular}{|c|c|c|c|c|} 
\hline 
Solution &$\Delta m^2$ & $B_{\perp {max}}$ & $\chi^2_{min}$ (Prob \%)&\\ 
\hline \hline 
$MHD-RSF_a$ & $1.1\times 10^{-8}$ & 80 &25.7 (32)& this work\\  
$MHD-RSF_s$ & $1.1\times 10^{-8}$ & 75 &30.5 (14)& this work\\  
\hline 
\hline 
 &$\Delta m^2$ & $\sin^2(2\vartheta)$ & $\chi^2_{min}$ (Prob \%)& Ref.\\ 
\hline 
\hline 
$ SMA_a$&$5.2\times 10^{-6}$&$4.7\times 10^{-3}$&29.7 (16)&\protect\cite{two,dark}\\
LMA&$2.4\times 10^{-5}$&0.78&27.0 (26)&\protect\cite{two,dark}\\ 
LOW&$1.0\times 10^{-7}$&0.93&32.0 (10)&\protect\cite{two,dark}\\
$ SMA_s$&$5.2\times 10^{-6}$&$4.7\times 10^{-3}$&32.0 (10)&\protect\cite{two,four}\\
VAC & $4.4\times 10^{-10}$ & 0.9 & 34.3 (6)&\protect\cite{four}\\
\hline
\hline
no-osc & &  & 87.9 ($6 \times 10^{-7}$) & \protect\cite{two}\\
\hline
\end{tabular} 
\vglue .3cm 
\caption{Best fit points and the corresponding probabilities for  
different solutions to the solar neutrino problem. The top row 
corresponds to the MHD-RSF solution presented here.} 
\label{comp} 
\end{center}
\end{table} 
Clearly the MHD-RSF fits seem somewhat better (though not in a
statistically significant way) than those obtained for the MSW effect
~\cite{two} as well as just--so solutions~\cite{four}. Notice that in
table~\ref{comp} we have used the same common calibrated theoretical
procedures and statistical criteria. These results are for the case
where the BP98 Standard Solar Model is adopted. 
We have also investigated the effect of varying the $hep$ flux,
obtaining a $hep$ normalization factor of 13.5 to be compared with 12
for the SMA solution, 38 for the LMA solution and 15 for the VAC
solution.

\subsection{Future}

Having performed our global analysis of the recent solar neutrino data
within the framework of our MHD-RSF solution to the solar neutrino
problem, we are in a position to calculate also the expected values of
a number of observables to be measured by future solar neutrino
experiments, such as SNO or Borexino.  This task has been developed
for the case of oscillation--type solutions to the solar neutrino
problem in ref.~\cite{Bahcall:2000zg}.  Here we will consider our
alternative MHD-RSF solution described in sections 2 and 3, because of
its theoretical elegance and the good quality of the global fits it
provides.  Again, the results of refs.~\cite{two,dark,four} will allow
us to compare quantitatively our simplest MHD-RSF predictions with
those associated with the vacuum (VAC) and MSW--type (SMA, LMA and
LOW) solutions to the solar neutrino problem using the same
well--calibrated theoretical calculation and fit procedure.

We determine the expected solar neutrino rates at SNO using the cross
sections for the CC and NC $\nu d$ reactions given by
ref.~\cite{Kubodera} and the preliminary SNO collaboration estimates
for the energy resolution, absolute energy scale and detection
efficiencies. For definiteness we adopt the most optimistic threshold
energy of 5 MeV which should be reached by the
collaboration~\cite{SNO}. We perform these calculations at the
best--fit points which we have determined in the present paper, using
90 and 99 \% CL error bars.  For definiteness we have considered the
global best fit points and local minima for $B_{\perp {max}}<100 kG$
given in table~\ref{global}, for the case $k_M=6$ and $r_0=0.6$ and
active-active MHD-RSF conversions.

We have calculated the neutral-to-charged-current event ratio (NC/CC
for short) and our results are presented in \fig{figsno1}.
\begin{figure}
  \centerline{\protect\hbox{\psfig
      {file=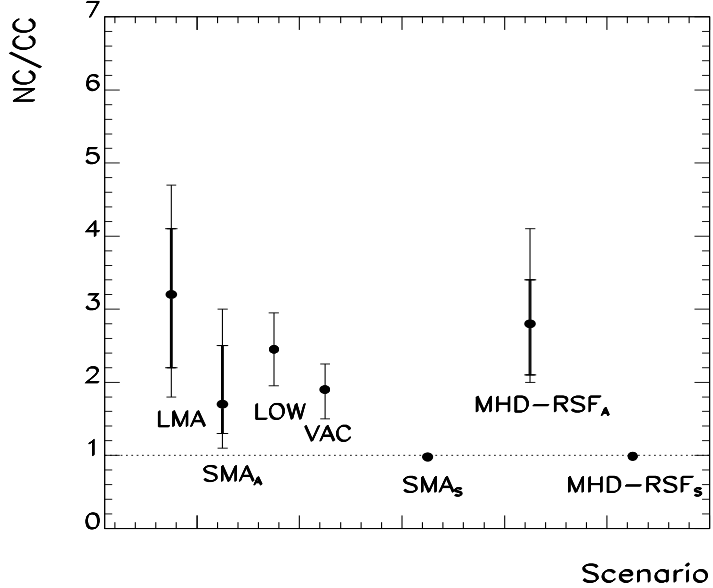,height=9cm,width=13cm,angle=0}}}
\caption{Neutral-to-charged-current event ratio expected at SNO for
  different solutions to the solar neutrino problem at 90\% CL and
  99\% CL. The no-oscillation or SM case is denoted by the horizontal
  line at one.}
\label{figsno1}
\end{figure}
Our predictions for the oscillation solutions agree relatively well
with those of \cite{Bahcall:2000zg}. The agreement is not perfect
because we use the full zenith angle dependence in the analysis of the
solar neutrino data instead of simply the day--night asymmetry
employed in ref.~\cite{Bahcall:2000zg}. The size of the error bars
displayed in \fig{figsno1} arises from the variation in the values of
neutrino oscillation parameters, rather than from statistical and
theoretical uncertainties, which are negligible~\cite{Bahcall:2000zg}.

Clearly from \fig{figsno1} we see that there is a substantial overlap
between our MHD-RSF predictions and those found for each of the
oscillation solutions (SMA, LMA, LOW, VAC). The overlap is especially
large between the LMA and the MHD-RSF solutions.
Taking into account the present theoretical uncertainties and a
reasonable estimate of the experimental errors attainable, it follows
that an unambiguous discrimination between our MHD-RSF solution and
the neutrino oscillation--type solutions to the solar neutrino problem
on the basis of the averaged event rates seems rather difficult.  The
expected features of the MHD-RSF recoil electron spectrum will be
discussed elsewhere~\cite{inprep}.

\section{Discussion \& Conclusions}

We have re-analysed the status of resonant spin--flavour solutions to
the solar neutrino problem in the framework of analytic solutions to
the solar magneto-hydrodynamics equations, using the most recent
global set of solar neutrino data.  We have shown that our procedure
is quite robust in the sense that the arbitrariness associated to the
magnetic field profile has been almost eliminated due to both
mathematical consistency and physical requirements.
Effectively our analysis substitutes neutrino mixing by a single
parameter $B_{\perp max}$ characterizing the maximum magnitude of the
magnetic field inside the convective region.
The value of $k_M$ characterizing the maximum number of individual
modes combined in a realistic profile and the parameter
$r_0$ characterizing the location of the convective region are
severely restricted. The allowed $k_M$ values are restricted by ohmic
dissipation arguments to be lower than 10 or so, and we have found
that our solar neutrino fits are pretty stable as long as $k_M$
exceeds 5.  Moreover our fits are pretty stable within the relevant
narrow range for $r_0$.
This way we obtain effective two--parameter global fits of solar
neutrino data for static MHD solutions characterized by $\Delta m^2$
and $B_{\perp max}$, since the magnetic field profile is essentially
unique. 
This enables us to compare their quality with that of the fits
obtained for the favored neutrino oscillation solutions to the solar
neutrino problem.  Adopting the Standard Solar Model we have found the
MHD-RSF fits to be slightly better than the oscillation fits, though
not in a statistically significant way.
We have also analysed the prospects to distinguish our best MHD-RSF
solution from the oscillation solutions (SMA, LMA, LOW, VAC) at future
solar neutrino experiments.
Both in the comparison of the present status of different solutions of
the solar neutrino problem, as well as in their future predictions at
SNO we have used a common well-calibrated theoretical procedure and
statistical criteria.
Taking into account the present theoretical uncertainties and the
expected experimental errors attainable, an unambiguous discrimination
between our MHD-RSF solution and the neutrino oscillation--type
solutions to the solar neutrino problem at the SNO experiment seems
rather difficult.
On the other hand better measurements of rate--independent solar
neutrino observables such as the day--night asymmetry and seasonality
would be potentially useful, since our MHD-RSF predictions differ from
the expectations of the oscillation schemes. For example, seasonality
is expected to be smaller~\cite{inprep} in our MHD-RSF solution than
in MSW~\cite{deHolanda:1999ty} or just--so oscillations ~\cite{four}.
On the other hand the day--night asymmetry of the MHD-RSF solution is
negligible, in contrast with the MSW solutions~\cite{inprep}.

Note, however, that the \emph{complete} MHD-RSF solution is
characterized also by a non--zero neutrino flavour mixing. This gives
it the potential to be discriminated from the oscillation--type
solutions~\cite{inprep}. The most distinctive signal expected in this
case consists of solar anti-neutrinos, which would provide a clear
signal in water Cerenkov experiments~\cite{Fiorentini:1998qe}.
Moreover, for large enough neutrino mixing one expects also a sizeable
suppression of the rates for $pp$ neutrinos, potentially testable at
the GNO experiment.
Last but not least, the possible time dependence of the charged
current signal due to solar cycles still remains as a possible tool to
discriminate the MHD-model from the oscillation schemes.

Note added: As we finished our paper there appeared the paper
E.~K.~Akhmedov and J.~Pulido, hep-ph/0005173, which also considers
predictions for SNO observables in the RSF scheme employing the
phenomenological magnetic field profiles they used previously in
ref~\cite{ad-hoc-profile2}.

\newpage
\appendix

\section{Zenith and Spectrum Data Samples and Fit Procedures}
\label{dataspec}

Here we summarize here the data used and the fit procedures adopted in
this paper.

The zenith dependence data given by the Super--Kamiokande
collaboration~\cite{sk99} are shown in table~\ref{datazenith}.
\begin{table}
\begin{center}
\begin{tabular}{|c|c|}
\hline
Angular Range & Data$_i\pm \sigma_{i}$ \\
\hline
Day $ 0<\cos\theta<1 $  & $0.463 \pm 0.0115$ \\
N1  $ -0.2<\cos\theta<0 $ & $0.512 \pm 0.026$\\
N2  $ -0.4<\cos\theta<-0.2 $ & $0.471 \pm 0.025$\\
N3  $ -0.6<\cos\theta<-0.4 $ & $0.506 \pm 0.021$\\
N4  $ -0.8<\cos\theta<-0.6 $ & $0.484 \pm 0.023$\\
N5  $ -1<\cos\theta<-0.8 $ & $0.478 \pm 0.023$\\
\hline
\end{tabular}
\caption{Super--Kamiokande Collaboration zenith angle distribution 
of events \protect\cite{sk99}.}
\label{datazenith}
\end{center}
\end{table}

The recoil electron spectrum data are given as
\begin{table}
\begin{center}
\begin{tabular}{|c|c|c|c|c|}
\hline
Energy bin & Data$_i \pm \sigma_{i,stat}$ & 
$\sigma_{i,exp}$ (\%) & $\sigma_{i,cal}$ (\%) & $\sigma_{i,uncorr}$ (\%) \\
\hline
    5.5 MeV $< E_e <6 $ MeV & $0.472 \pm 0.037$ 
    &  1.3 &  0.3  & 4.0   \\
    6 MeV $< E_e <6.5 $ MeV & $0.444 \pm 0.025$ 
    &  1.3 &  0.3  & 2.5  \\
    6.5 MeV $< E_e <7 $ MeV & $0.427 \pm 0.022$ 
    & 1.3  & 0.3  & 1.7  \\
    7 MeV $< E_e <7.5 $ MeV & $0.469 \pm 0.022$ 
    & 1.3  & 0.5  & 1.7  \\
    7.5 MeV $< E_e <8 $ MeV & $0.516 \pm 0.022$ 
    & 1.5  & 0.7  & 1.7  \\
    8 MeV $< E_e <8.5 $ MeV & $0.488 \pm 0.025$ 
    & 1.8  & 0.9  & 1.7  \\
    8.5 MeV $< E_e <9 $ MeV & $0.444 \pm 0.025$ 
    & 2.2  & 1.1   & 1.7 \\
    9 MeV $< E_e <9.5 $ MeV & $0.454 \pm 0.025$ 
    & 2.5  &  1.4 & 1.7  \\
    9.5 MeV $< E_e <10 $ MeV & $0.516 \pm 0.029$ 
    & 2.9   & 1.7  & 1.7  \\
    10 MeV $< E_e <10.5 $ MeV & $0.437 \pm 0.030$ 
    &3.3   & 2.0  & 1.7  \\
    10.5 MeV $< E_e <11 $ MeV & $0.439 \pm 0.032$ 
    &3.8   & 2.3  & 1.7   \\
    11 MeV $< E_e <11.5 $ MeV & $0.476 \pm 0.035$ 
    &4.3   & 2.6  & 1.7  \\
    11.5 MeV $< E_e <12 $ MeV & $0.481 \pm 0.039$ 
    &4.8   & 3.0  & 1.7   \\
    12. MeV $< E_e <12.5 $ MeV & $0.499 \pm 0.044$ 
    &5.3   & 3.4  & 1.7  \\
    12.5 MeV $< E_e <13 $ MeV & $0.538 \pm 0.054$ 
    & 6.0  & 3.8  & 1.7   \\
    13 MeV $< E_e <13.5 $ MeV & $0.530 \pm 0.069$ 
    & 6.6  & 4.3  & 1.7  \\
    13.5 MeV $< E_e <14 $ MeV & $0.689 \pm 0.092$ 
    &  7.3 & 4.7  & 1.7  \\
    14 MeV $< E_e<20  $ MeV  & $0.612 \pm 0.077$ 
    & 9.2  & 5.8  &  1.7 \\ 
\hline
\end{tabular}
\caption{Recoil energy spectrum of solar neutrinos 
from the 825-day Super--Kamiokande Collaboration data sample
\protect\cite{sk99}. 
}
\label{spectrum}
\end{center}
\end{table}
In table~\ref{spectrum} $\sigma_{i,stat}$ is the statistical error,
$\sigma_{i,exp}$ is the error due to correlated experimental errors,
$\sigma_{i,cal}$ is the error due to the calculation of the expected
spectrum, and $\sigma_{i,uncorr}$ is due to uncorrelated systematic
errors.

In our study we use the experimental results from the
Super--Kamiokande Collaboration on the recoil electron spectrum on the
18 energy bins including the results from the LE analysis for the 16
bins above 6.5 MeV and the results from the SLE analysis for the two
low energy bins below 6.5 MeV, shown in table~\ref{spectrum}.

Notice that in table~\ref{spectrum} we have symmetrized the errors to
be included in our $\chi^2$ analysis. We have explicitly checked that
the exclusion region is very insensitive to this symmetrization.  We
define $\chi^2$ for the spectrum as
\begin{equation}
\chi^2_S=\sum_{i,j=1,18} 
(\alpha_{sp}\frac{\displaystyle R^{th}_i}{R^{\rm BP98}_i} 
-R^{exp}_i)
\sigma_{ij}^{-2} (\alpha_{sp}\frac{\displaystyle R^{th}_j}{R^{\rm BP98}_j}
 - R^{exp}_j)
\end{equation}
where
\begin{equation}
\sigma^2_{ij}=\delta_{ij}(\sigma^2_{i,stat}+\sigma^2_{i,uncorr})+
\sigma_{i,exp} \sigma_{j,exp}+\sigma_{i,cal}\sigma_{j,cal}
\end{equation}
Again, we introduce a normalization factor $\alpha_{sp}$ in order to
avoid double-counting with the data on the total event rate which is
already included in $\chi^2_R$.  Notice that in our definition of
$\chi^2_S$ we introduce the correlations amongst the different
systematic errors in the form of a non-diagonal error matrix in
analogy to our previous analysis of the total rates.  These
correlations take into account the systematic uncertainties related to
the absolute energy scale and energy resolution.

The general expression of the expected rate in the presence of
oscillations $R^{th}$ in a bin, is given from Eq.(\ref{ratesth}) but
integrating within the corresponding electron recoil energy bin and
taking into account that the finite energy resolution implies that the
{\em measured } kinetic energy $T$ of the scattered electron is
distributed around the {\em true } kinetic energy $T'$ according to a
resolution function $Res(T,\,T')$ of the form 
\begin{equation}
Res(T,\,T') = \frac{1}{\sqrt{2\pi}s}\exp\left[
{-\frac{(T-T')^2}{2 s^2}}\right]\ ,
\end{equation}
where
\begin{equation}
s = s_0\sqrt{T'/{\rm MeV}}\ ,
\label{Delta}
\end{equation}
and $s_0=0.47$ MeV for Super--Kamiokande \cite{sk800,Faid}. On the
other hand, the distribution of the true kinetic energy $T'$ for an
interacting neutrino of energy $E_\nu$ is dictated by the differential
cross section $d\sigma_\alpha(E_\nu,\,T')/dT'$, that we take from
\cite{CrSe}. The kinematic limits are
\begin{equation}
0\leq T' \leq {\overline T}'(E_\nu)\ , 
\ \ {\overline T}'(E_\nu)=\frac{E_\nu}{1+m_e/2E_\nu}\ .
\end{equation}
For assigned values of $s_0$, $T_{\rm min}$, and $T_{\rm max}$, the
corrected cross section $\sigma_{\alpha}(E)$ $(\alpha = e,\,x)$ is
given as
\begin{equation}
\sigma_{\alpha}(E_\nu)=\int_{T_{\rm min}}^{T_{\rm max}}\!dT
\int_0^{{\overline T}'(E_\nu)}
\!dT'\,Res(T,\,T')\,\frac{d\sigma_{\alpha}(E_\nu,\,T')}{dT'}\ .
\label{sigma}
\end{equation}

\vskip1cm

{\Large \bf Acknowledgements}\\

We thank Alexei Bykov, Vladimir Kutvitsky, Dmitri Sokoloff and Victor
Popov for useful discussions.
This work was supported by DGICYT grant PB98-0693, by the European 
Commission under Intas Project 96-0659 and TMR contract ERBFMRX-CT96-0090, 
and by an Iberdrola research excellence grant.  VBS and TIR were 
partially supported by the RFBR grant 00-02-16271, CPG was supported 
by the Generalitat Valenciana grant GV99-3-1-01 and OGM was 
supported by the CONACyT-Mexico grant J32220-E.

\end{document}